\documentclass[prl,twocolumn,showpacs,preprintnumbers,amsmath,amssymb]{revtex4}%

\usepackage{graphicx}
\usepackage{dcolumn}
\usepackage{bm}

\begin{document}


\title{Zoomable telescope by rotation of toroidal lenses}

\author{Stefan Bernet}

\affiliation{Division of Biomedical Physics, Medical University of Innsbruck, M{\"u}llerstr. 44, A-6020 Innsbruck, Austria}%


\date{\today}

\begin{abstract}
A novel type of a continuously zoomable telescope is based on two pairs of adjacent toroidal lenses ("saddle lenses") in combination with standard optical components. Its variable magnification is adjusted by a mere rotation of the four saddle lenses around the optical axis. This avoids the necessity of classical zoom systems to shift multiple lenses along the longitudinal axis of the setup. A rotationally tunable pair of saddle lens consists of two individual saddle lenses (also known as quadrupole lenses, or biconic lenses), which are arranged directly behind each other, acting as a  "combi-saddle lens". The transmission function of such a combi-saddle lens corresponds to that of a single saddle lens, but with an adjustable optical power which depends on the mutual rotation angle between its two components. The optical system contains two of these combi-saddle lenses, and acts as a cylindrical Kepler telescope in one plane, and as a cylindrical Galilei telescope in the orthogonal plane. The two orthogonal Kepler/Galilei telescopes stay aligned and change their magnification factors in the same way when the telescope is zoomed by adjusting the optical powers of the two combi-saddle lenses. Altogether this produces a sharp image which is mirrored with respect to the axis of the Kepler telescope. Thus, in addition to the zooming capabilities of the telescope, it is also possible to rotate the resulting image by a rotation of the whole telescope, or of all included saddle lenses. We theoretically explain the operation principle of the telescope in both, a ray-optical and a wave-optical description.  
\end{abstract}

\maketitle


\section{Introduction}
\begin{figure}[ht]
\begin{center}
\includegraphics[width=\columnwidth]{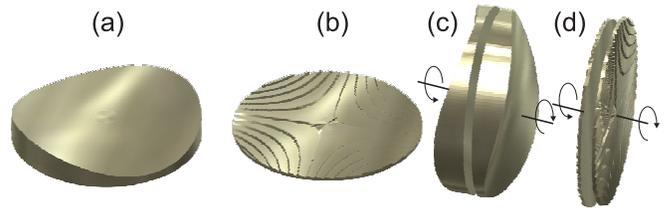}
\end{center}
\caption[]{(a): Shape of a refractive saddle lens. (b): Corresponding "thin" diffractive saddle lens, obtained from a modulo operation from (a). (c): Refractive "combi-saddle lens" consisting of two stacked elements shown in (a). The combination corresponds to another saddle lens, with an optical power which is tunable by a relative rotation between its two components (indicated in the figure). (d): Diffractive version of (c), corresponding to a thin diffractive "combi-saddle lens". Note that the coloration of the lenses is used for visualization only, - actually  the elements are transparent.}
\label{stl}
\end{figure}

Anamorphic optical setups, i.e. arrangements with no circular symmetry around the optical axis, can be used for various tasks in beam manipulation, like beam (or image) rotation, or asymmetric image scaling, i.e. optically transforming the aspect ratio of an image \cite{Braunecker, Davis, Lohmann, Wang, Iemmi, Wolf}. The corresponding image transformations are typically described in the ray optics regime, using four component vectors (for the positions and directions in the $x$- and $y$-planes, respectively) propagated by a generalized version of matrix optics using $4 \times 4$ matrices \cite{Lohmann, Moreno, Wolf, Rodrigo}. Such anamorphic optical setups may be implemented with combinations of cylindrical lenses with different orientations. However, a greater flexibility is possible if anamorphic freeform optical elements are used, like rotationally asymmetric  toroidal (or biconic) lenses. Toroidal lenses are anamorphic optical elements, which primarily have a transmission function corresponding to that of two crossed cylindrical lenses, which generally may have different focal lengths. A special class of them, consisting of two orthogonal cylindrical lenses with opposite optical power, are also known as  "saddle lenses" \cite{Sales}, or optical quadrupol lenses \cite{Weber}, since their effect is analogous to that of electric or magnetic quadrupol lenses used in electron optics \cite{Hawkes}. For reasons of simplicity we speak of a saddle lens with a quadrupole focal length of $f$, or a corresponding quadrupole optical power of $f^{-1}$, if its focal lengths along its two orthogonal main axes is $+f$ and $-f$, respectively. An example for a saddle lens is sketched in Fig. \ref{stl}(a). A diffractive version of such an element can be obtained by wrapping the phase function of element (a) by a modulo operation into a thin, blazed surface profile, as indicated in Fig. \ref{stl}(b). An interesting property of saddle lenses is that two of them can be stacked, i.e. mounted directly behind each other, and (in a thin lens approximation) the corresponding "combi-saddle lens" has again the transmission function of a single saddle lens, however with a different quadrupole optical power which can be tuned by rotating one element with respect to the other around the optical axis. This is indicated in Fig.  \ref{stl}(c) and (d) for refractive and diffractive versions of combined saddle lenses, respectively.

The insertion of two combi-saddle lenses in  certain optical setups allows one to construct a zoom system, which basically acts as an afocal telescope, whose angular magnification can be continuously tuned by a rotation of the individual saddle lens elements around the optical axis. This feature will be discussed in both, a ray-optical and a wave-optical description. In the ray-optical approach it turns out that the telescope actually acts as a combination of a cylindral Kepler telescope with a cylindrical Galilei telescope in two orthogonal planes, which both stay aligned and yield the same zoom factor when tuning the magnification by a rotation of the saddle lenses.  In the wave-optical approach it will be shown that the working principle of a saddle lens telescope is based on the interesting fact that an (optically realizable) convolution of an input image with the transmission function of a saddle lens just yields  a Fourier transform of the input image, which is however scaled, depending on the adjusted quadrupole optical power of the saddle lens. Additionally the result of the convolution has an offset phase modulation, which corresponds to the phase of another saddle lens. However, this offset phase modulation can be compensated in the telescope by inserting a corresponding conjugate saddle lens in the respective Fourier plane. Then a subsequent optically performed Fourier transformation yields the scaled output image. In order to be able to optically perform the initial convolution between the input image and a saddle lens, an additional property of saddle lenses is used, namely that their Fourier transform still corresponds to the transmission function of a saddle lens, however with a different quadrupole optical power. 

\section{Tunable combined saddle lenses}

The transmission function of a saddle lens with a quadrupole optical power of $f_s^{-1}$ corresponds to that of two crossed cylindrical lenses with equal positive and negative optical powers of $\pm f_s$ along the $y-$ and $x-$axes. In the following  it is denoted as $S(x,y,f_s^{-1})$, where the third argument corresponds to the quadrupole optical power of the saddle lens:

\begin{equation}\label{equ:saddle}
S(x,y,f_s^{-1})=\exp[i \frac{\pi}{\lambda f_s} (x^2-y^2)].
\end{equation}

A \emph{tunable} saddle lens can be constructed by different methods. One of them is to just stack two individual saddle lenses with equal quadrupole optical powers into a single, combined element, a "combi-saddle lens", as shown in Fig. \ref{stl}(c) or (d). The combination of two saddle lenses with transmission functions $S_{\alpha}$ and  $S_{-\alpha}$ which are rotated by an angle $\alpha$ in positive and negative directions, respectively (such that the total rotation angle between the two elements is $2 |\alpha|$), produces a resulting transmission function of the combi-saddle lens $T_{\mathrm{C}}$ given by:

\begin{eqnarray}\label{equ:combisaddle}
T_{\mathrm{C}}=S_{\alpha}(f_s^{-1}) S_{-\alpha}(f_s^{-1}) &=& \exp[i  \frac{\pi}{\lambda f_s}  2\cos(2 \alpha) (x^2-y^2)] \nonumber \\ &=& S(x,y,2 f_s^{-1} \cos(2 \alpha)). 
\end{eqnarray}

This transmission function corresponds again to that of a single saddle lens with the same orientation of its main axes, but with a different quadrupole optical power of:
\begin{equation}\label{equ:factm1}
f_{\mathrm{c}}^{-1}= 2 f_s^{-1} \cos(2 \alpha).
\end{equation}

A second method to realize a tunable saddle lens is to just combine two cylindrical lenses with opposite optical powers in a "scissor-" arrangement \cite{Braunecker, Wolf}, i.e. with a variable angle between the cylinder axes. For example, if we assume a combination of a positive and a negative cylindrical lens with respective optical powers of $\pm f_{cyl}$, one of them oriented parallel to the $x-$ axis, and the other parallel to the $y-$axis, then an additional rotation of the two lenses in opposite directions by an angle of $\pm \alpha$ results in the combined transmission function:

\begin{equation}\label{equ:combicyl}
T_{\mathrm{C}}=\exp[i \frac{\pi}{\lambda f_{cyl}} \cos(2 \alpha)(x^2-y^2)] = S(x,y,f_{cyl}^{-1} \cos(2 \alpha)),
\end{equation}
which again corresponds to a saddle lens. However, its tuning range is only half of that of  two combined saddle lenses with equal optical power (Eq. \ref{equ:combisaddle}).  More generally, any combination of two mutually rotatable cylindrical (or toroidal) lenses in combination with adequately chosen spherical lenses (which correct for additionally appearing spherical lens terms) can be used as a tunable combi-saddle lens. However, the construction of a combi-saddle lens using two individual saddle lenses of equal optical power maximizes the tuning range for the optical power of the combined element, and thus also reduces optical aberrations as compared to other combinations of toroidal lenses which yield the same tuning range.

\section{Ray-optical description of a tunable saddle lens telescope}
\begin{figure}[t]
\begin{center}
\includegraphics[width=\columnwidth]{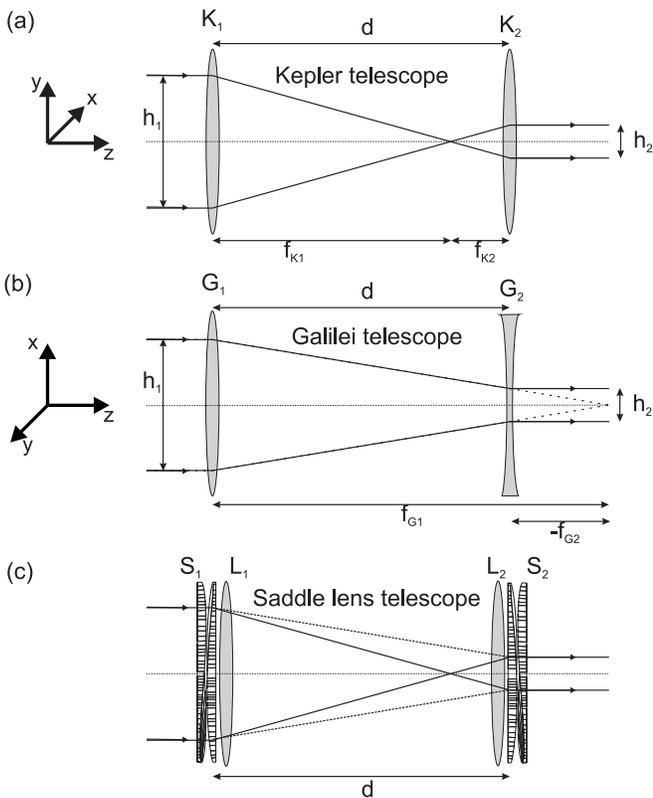}
\end{center}
\caption[]{(a): Layout of a cylindrical Kepler telescope in the (y,z)-plane. (b): Layout of a cylindrical Galilei telescope in the orthogonal (x,z)-plane. (c): Combination of a cylindrical Kepler telescope and a cylindrical Galilei telescope in two orthogonal planes within a tunable saddle lens telescope.}
\label{classical}
\end{figure}

In order to explain the principle of a saddle lens telescope, which combines a Kepler and a Galilei telescope in two orthogonal planes, we first recapitulate the optical layouts for Kepler and Galilei telescopes, as shown in Fig. \ref{classical}(a) and (b), respectively. Figure \ref{classical}(a) shows the beam path in a Kepler telescope in the (y,z)-pane, consisting of two positive cylindrical lenses K$_1$ and K$_2$ at a distance $d$, with respective focal lengths of $f_{K1}$ and $f_{K2}$, respectively. The telescope is aligned, if the back focal plane of the first lens K$_1$ coincides with the front focal plane of K$_2$, i.e. if:
\begin{equation}\label{equ:Kepler1}
f_{K1}+f_{K2}=d.
\end{equation}
The angular magnification $m$ of the telescope is inversely proportional to the reduction of the beam diameter of a parallel incident beam with diameter $h_1$, which is transformed into a parallel outgoing beam with diameter $h_2$, i.e.:
\begin{equation}\label{equ:Kepler2}
m=\frac{h_1}{h_2}=\frac{f_{K1}}{f_{K2}}.
\end{equation}
For a given distance $d$ between the two lenses, and a given magnification factor $m$, the required optical powers of the two lenses can be easily calculated as:
\begin{equation}\label{equ:Kepler3}
f_{K1}^{-1}=d^{-1}+\frac{d^{-1}}{m}  \;\;  \mathrm{and}  \;\; f_{K2}^{-1}=d^{-1}+m d^{-1}.
\end{equation}

Figure \ref{classical}(a)  shows a cylindrical Galilei telescope in the orthogonal (x,z)-plane consisting of a positive cylindrical lens G$_1$ with focal length $f_{G1}$, and a negative (concave) cylindrical lens G$_2$ with the (negative) focal length $f_{G2}$. Both lenses have the same distance $d$, as within the Kepler telescope in (a). The telescope is aligned if:
\begin{equation}\label{equ:Galilei1}
f_{G1}+f_{G2}=d.
\end{equation}
Note, however, that $f_{G2}$ is now negative. The angular magnification $m$ of the Galilean telescope should correspond to that of the Keplerian telescope in (a), and is given by:
\begin{equation}\label{equ:Galilei2}
m=\frac{h_1}{h_2}=-\frac{f_{G1}}{f_{G2}}.
\end{equation}
Thus, for a given magnification $m$, and a given distance $d$ between the two lenses, the required optical powers of the two lenses are:
\begin{equation}\label{equ:Galilei3}
f_{G1}^{-1}==d^{-1}-\frac{d^{-1}}{m}   \;\;  \mathrm{and}  \;\; f_{G2}^{-1}=d^{-1}-m d^{-1}.
\end{equation}
Comparing Eqs. \ref{equ:Kepler3} and \ref{equ:Galilei3} it is obvious that one can replace the orthogonal cylindrical lenses K$_1$ and G$_1$ of the Kepler and the Galilei telescopes in a combined telescope by a combination of a rotationally symmetric lens L$_1$ with a successive (combi-)saddle lens S$_1$, whose respective optical powers are:
\begin{equation}\label{equ:Saddle1}
f_{L1}^{-1}=d^{-1}	 \;\;		\mathrm{and} \;\; 		f_{S1}^{-1}=\pm \frac{d^{-1}}{m}.	
\end{equation}
It can easily be checked that (in a thin lens approximation) the mutual transmission function of these two subsequent lenses corresponds to that of $K_1$ and $G_1$ in the two orthogonal planes of the cylindrical Kepler and the Galilei telescopes, respectively.

Analogously, one can combine the two orthogonal cylindrical lenses K$_2$ and G$_2$ into a set of a second rotationally symmetric lens L$_2$, and a second saddle lens S$_2$ with corresponding optical powers of:
\begin{equation}\label{equ:Saddle2}
f_{L2}^{-1}=d^{-1}	 \;\;		\mathrm{and} \;\; 		f_{S2}^{-1}=\pm m d^{-1}.	
\end{equation}
For all magnification factors $m$ the combinations consist of the same (i.e. static) rotationally symmetric lenses (L$_1$ and L$_2$), which both have the same optical power of $d^{-1}$, and of two saddle lenses S$_1$ and S$_2$, with a variable optical power which depends on the magnification factor. Thus, if these saddle lenses are implemented by combi-saddle lenses, whose optical powers are adjusted to correspond to $f_{S1}^{-1}$ and $f_{S2}^{-1}$ in Eqs. \ref{equ:Saddle1} and \ref{equ:Saddle2}   respectively, it is possible to construct a combined Kepleri/Galilei telescope (in orthogonal axes), with an adjustable magnification factor, which is controlled purely by the adjustment of the two combi-saddle lenses. 

Such a "saddle lens telescope" which combines the two orthogonal cylindrical telescopes of (a) and (b) is shown in Fig. \ref{classical}(c). There, L$_1$ and L$_2$ are two rotationally symmetric lenses, both with the same optical power of $f_{L1}^{-1}=f_{L2}^{-1}=d^{-1}$, whereas S$_1$ and S$_2$ are two saddle lenses with optical powers of $f_{S1}^{-1}=d^{-1}/m$ and $f_{S2}^{-1}=m d^{-1}$, respectively, and $m$ is the magnification of the telescope. The main axes of the two saddle lenses S$_1$ and S$_2$ are oriented parallel. The beam paths in the yz-plane (solid lines) and the xz-plane (dashed lines) correspond to that of the Kepler telescope in (a), and the Galilei telescope in (b), respectively. 

The combined Kepler/Galilei telescope inverts an incoming beam along the axis of the Kepler telescope, since there is a focus within the beam path, whereas the beam is not inverted along the axis of the Galileian telescope. If such a telescope is used for imaging, this results in a mirroring operation with respect to the axis of the Kepler telescope. Thus a rotation of the whole saddle lens telescope (or just a rotation of all included combi-saddle lenses) results in an image rotation of the mirror-inverted image by twice the rotation angle of the telescope. For correct operation of the combined telescope the main axes of the two combi-saddle lenses have to stay parallel. This is achieved if the magnification is tuned by rotating the two elements of each combi-saddle lens always symmetrically (by the same angle) into opposite directions. The corresponding rotation angles for a certain magnification factor can be calculated from Eqs \ref{equ:Saddle1} and \ref{equ:Saddle2}, together with Eq. \ref{equ:factm1}.

\begin{figure}[t]
\begin{center}
\includegraphics[width=\columnwidth]{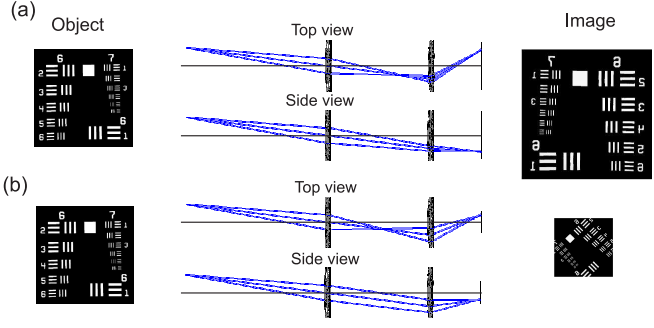}
\end{center}
\caption[]{Ray-optical simulation of an imaging setup based on the layout of Fig.\ref{classical}(c) using commercial ray-tracing software (Zemax Optics Studio). For this purpose the afocal telescope of Fig.\ref{classical}(c) is expanded with an additional focusing lens in front of the first saddle lens, and an additional objective lens behind the second combi-saddle lens. (a) shows the top view and the side view of the beam paths for a magnification factor of $m=1.33$. The original object and the simulated image are shown at the left, and at the right sides, respectively. Note that the output image is mirrored with respect to the input image. (b) shows a second simulation with the magnification factor $m=0.66$. Additionally all combi-saddle lenses are counter-clockwise rotated by an angle of 22.5$^{\circ}$, resulting in a counter-clockwise rotation of the (mirrored) image by 45$^{\circ}$.}
\label{zemax}
\end{figure}

A ray-optical simulation of the imaging properties of a setup based on the layout of  Fig. \ref{classical}(c) is shown in Fig.  \ref{zemax}. Since the original setup of Fig. \ref{classical}(c) is afocal (acting between infinitely distant object and image planes), the setup in the simulation was expanded by an additional focusing lens directly in front of the telescope at the object side, and an additional objective lens directly behind the telescope at the image side. The simulations were performed using a test image (left side) which was magnified and demagnified by the factors $m=1.33$ and $m=0.66$ in (a) and (b), respectively. The spherical lenses and the (rotatable) combi-saddle lenses were simulated by refractive (bulk) elements, and geometric lens errors were included in the simulation. Additionally all saddle lenses in (b) were rotated counter-clockwise by an angle of 22.5$^{\circ}$, which results in an image rotation of 45$^{\circ}$, as expected from our previous discussion. The simulations also confirm that the resulting image is a mirrored version of the input image, as expected.

The optical setup of Fig. \ref{classical}(c) with the focal lengths of the two lenses L$_1$ and L$_2$ corresponding to their distance, i.e. $f_{L1}=f_{L2}=d$, is a special optical arrangement which is known \cite{Lohmann} to perform a mutual optical Fourier transform between the planes of the two combi-saddle lenses S$_1$ and S$_2$. Actually the condition that the two saddle lenses are located in mutual Fourier planes is required for the construction of a saddle lens telescope. However, since an optical Fourier transform can also be performed in another setup, this suggests a further layout of a saddle lens telescope, which will be described in the following wave-optical discussion.

\section{Wave-optical description}
An important property of saddle lenses is that a convolution of a two-dimensional master image $g(x,y)$ with a combi-saddle lens $T_{\mathrm{C}}$ (according to Eq. \ref{equ:combisaddle}), leads to a convolved image $g^*(u,v)$ given by

\begin{eqnarray}\label{equ:conv}
g^*(u,v)&=&g(x,y) \circledast \exp[i \frac{m \pi}{\lambda f_s} (x^2-y^2)] \\  &=&\int g(x,y)\exp(i \frac{m \pi}{\lambda f_s} [(u-x)^2-(v-y)^2]) \, \mathrm{d}x \mathrm{d}y, \nonumber
\end{eqnarray}
where '$\circledast$' denotes the convolution symbol, and $m$ is the adjustable scale factor for the optical power of the saddle lens.
Processing of this convolution integral yields:

\begin{eqnarray}\label{equ:conv1}
&& g^*(u,v)=\exp(i \frac{m \pi}{\lambda f_s} [u^2-v^2]) \times \\ &&\int g(x,y)\exp(i \frac{m \pi}{\lambda f_s} [x^2-y^2]) \exp\left[-i \frac{m 2 \pi}{\lambda f_s} (u x - vy)\right] \, \mathrm{d}x \mathrm{d}y. \nonumber
\end{eqnarray}
 
The integral actually corresponds to a Fourier transform $\mathcal{F} \{...\}(u,v)$ of the function $g(x,y) \exp(i \pi m /(\lambda f_s) [x^2-y^2])$,  i.e.:

\begin{eqnarray}\label{equ:conv2}
&& g^*(u,v)=\exp(i \frac{m \pi}{\lambda f_s} [u^2-v^2]) \times \\ && \mathcal{F}\left\{g(x,y)\exp(i\frac{m \pi}{\lambda f_s} [x^2-y^2])\right\}(m k_x,-m k_y), \nonumber
\end{eqnarray}
where
\begin{equation}\label{equ:kcoord}
k_x = \frac{2 \pi}{\lambda} \frac{u}{f_s}  \quad \mathrm{and} \quad k_y = \frac{2 \pi}{\lambda} \frac{v}{f_s}. 
\end{equation}

Note that the $(k_x, k_y)$-coordinates in Fourier space in Eq. (\ref{equ:conv2}) are scaled by the factor $m$,  and the $k_y$ coordinate is negative, which corresponds to a mirroring operation with respect to the $y-$axis. Due to the similarity theorem and the symmetry properties of the Fourier transform, this can be expressed in unscaled and non-mirrored coordinates $(k_x, k_y)$ by transferring the scaling and $y-$mirroring operations into the kernel of the Fourier transform, i.e.:
 
\begin{eqnarray}\label{equ:conv3}
&& g^*(u,v)= \exp(i \frac{m \pi}{\lambda f_s}  [u^2-v^2]) \times \\ &&  \mathcal{F}\left\{g^{\mathrm{R}}(\frac{x}{m},\frac{y}{m}) \exp(i \frac{\pi}{m \lambda f_s} [x^2- y^2])\right\}(k_x, k_y), \nonumber
\end{eqnarray}
where $g^{\mathrm{R}}(x/m,y/m)=g(x/m,-y/m)$ is the mirror image (with respect to the $y-$axis) of $g$.
 
Thus the convolution of an arbitrary 2-dimensional function $g(x,y)$ with a saddle lens basically corresponds to a Fourier transform of the (with respect to the $y-$axis) mirrored  function $g$, which is additionally scaled by the factor $m$, and multiplied by another saddle lens term. 
Since the $y-$axis is defined as the optical axis of the convex cylindrical sub-lens within the saddle lens, a rotation of the saddle lens by an angle $\theta$ rotates the $y-$axis, i.e. the axis of the mirroring operation, and thus rotates the Fourier transform of the resulting image by $2\theta$.

This convolution of the two-dimensional image $g(x,y)$ with the saddle lens $\exp(i \pi m /(\lambda f_s)  [x^2- y^2])$ can be also performed using the convolution theorem, i.e.:

\begin{eqnarray}\label{equ:convtheo}
g^*(u,v)&=&g(x,y) \circledast \exp(i\frac{m \pi}{\lambda f_s} [x^2-y^2]) \\ &=& \mathcal{F}^{-1}\{\mathcal{F}\{g(x,y)\} \cdot \mathcal{F}\{\exp(i\frac{m \pi}{\lambda f_s} [x^2-y^2])\}\}, \nonumber
\end{eqnarray}
where $\mathcal{F}^{-1}$ is the inverse Fourier transform. In this equation, the (optically implementable) Fourier transform of the input image has to be multiplied by the Fourier transform of a saddle lens. Fortunately, in an optical implementation this multiplication just corresponds to the transmission of the field through another saddle lens, since it interestingly turns out that the Fourier transform of a saddle lens is again a saddle lens, however with another optical power, namely
\begin{eqnarray}\label{equ:FTSaddle}
\mathcal{F}\{\exp(i\frac{m \pi}{\lambda f_s} [x^2-y^2])\} &=& \exp(-i \frac{\lambda f_s}{4 m \pi} [k_x^2- k_y^2]) \\ &=& \exp(-i \frac{\pi}{m \lambda f_s} [u^2- v^2]). \nonumber
\end{eqnarray}
Thus Eq. \ref{equ:convtheo} can be rewritten:
\begin{equation}\label{equ:convtheo1}
g^*(u,v)= \mathcal{F}^{-1}\{\mathcal{F}\{g(x,y)\} \cdot  \exp(-i \frac{\pi}{m \lambda f_s} [u^2- v^2]) \}.
\end{equation}
Using Eq. \ref{equ:conv3} one obtains:
\begin{eqnarray}\label{equ:result0}
 && \exp(i \frac{m \pi}{\lambda f_s}  [u^2-v^2])  \mathcal{F} \left\{ g^{\mathrm{R}}(\frac{x}{m},\frac{y}{m}) \exp(i \frac{\pi}{m \lambda f_s} [x^2- y^2])\right\} \nonumber \\ && =  \mathcal{F}^{-1} \left\{ \mathcal{F}\{g(x,y)\} \cdot  \exp(-i \frac{\pi}{m \lambda f_s} [u^2- v^2]) \right\}.
\end{eqnarray}
This Equation can be converted into: 
\begin{eqnarray}\label{equ:result1}
&& g^{\mathrm{R}}\left(\frac{x}{m},\frac{y}{m} \right) =\exp(-i \frac{\pi}{m \lambda f_s} [x^2- y^2]) \times \\ && \mathcal{F}^{-1} \Bigg\{ \exp(-i \frac{m \pi}{\lambda f_s} [u^2-v^2]) \times \nonumber \\ && \mathcal{F}^{-1} \left[ \mathcal{F}\{g(x,y)\} \cdot \exp(-i \frac{\pi}{m \lambda f_s}  [u^2- v^2]) \right] \Bigg\}. \nonumber
\end{eqnarray}
The left side of the equation ($g^{\mathrm{R}}\left(\frac{x}{m},\frac{y}{m} \right)$) corresponds to the zoomed (magnification factor $m$) and $y-$mirrored input image $g(x,y)$, whereas the right side corresponds to a sequence of operations consisting of Fourier transformations and multiplications with the adjustable transmission functions of saddle lenses, which all can be optically performed.

\section{Optical implementation}
\begin{figure}[t]
\begin{center}
\includegraphics[width=\columnwidth]{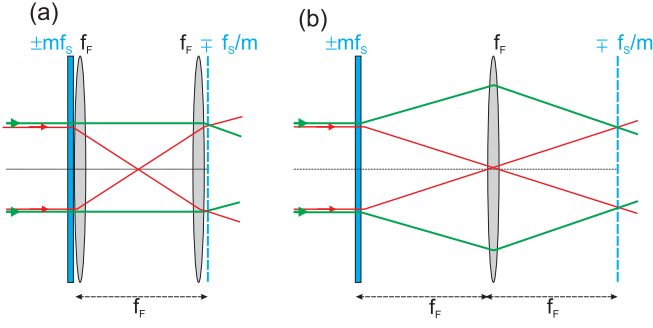}
\end{center}
\caption[]{Two methods to perform a Fourier transform of a saddle lens (blue): (a) The saddle lens (blue) with quadrupole focal length $\pm m f_s$ is illuminated with a plane wave. A spherical Fourier transforming lens (gray) with focal length $f_F$ located directly behind it performs a Fourier transform of the transmission function of the saddle lens at a distance $f_F$ (dashed blue plane), where a second spherical lens with focal length $f_F$ is located to correct for the phase.  (b): A spherical lens with focal length $f_F$ is located at a distance $f_F$ behind the saddle lens. At a further distance $f_F$ behind the spherical lens, the Fourier transform of the saddle lens is obtained (dashed blue). In both cases (a) and (b) the light field in the Fourier plane  (indicated by the dashed blue lines) has the phase profile of a conjugate saddle lens (with respect to the original saddle lens), with a quadrupole focal length of $\mp f_s/m$. The situation is sketched for $f_s=f_F$, and $m=1$, indicating the optical rays along the two perpendicular optical axes of the saddle lens in green and red, respectively. In this case the size of the Fourier transformed saddle lens corresponds to that of the initial saddle lens.}
\label{fft}
\end{figure}

Fourier transformations can be implemented optically in different setups \cite{Lohmann}. Figure \ref{fft}(a) shows a setup using two spherical lenses (gray) with the same focal lengths $f_F$ seperated by a distance $f_F$. 
In this case  an object (blue) located directly in front of the first lens is optically Fourier transformed in a plane directly behind the second lens (dashed blue plane). A second method to perform an optical Fourier transform is shown in Fig. \ref{fft}(b). There, a spherical lens with focal length $f_F$ is located between the two mutual Fourier planes, which correspond to the front and back focal planes of the lens, respectively.  In both cases (a) and (b), the field of the Fourier transformed object is identical. Generally a Fourier transform acts between two reciprocal coordinate systems, e.g. it transforms the Cartesian $(x,y)$-coordinates into $(k_x,k_y)$-coordinates in reciprocal (momentum) space, where $k_x=2\pi/x$ and $k_y=2\pi/y$. On the other hand the optically implemented Fourier operation (with lenses of focal lengths $f_F$) transforms a two-dimensional object in $(x,y)$-space into a Fourier transformed image in $(x,y)$-space, which is again Cartesian. In a paraxial approximation, the connection between the two coordinate systems is:
\begin{equation}\label{equ:trafo}
(x,y) =(\frac{k_x}{k_0} f_F,\frac{k_y}{k_0} f_F),
\end{equation} 
where $k_0=2\pi/\lambda$. Therefore the Cartesian $(u,v)$-coordinates in the plane of the optically performed convolution of the input image with a saddle lens, whose optical power is scaled in units of  $f_s^{-1}$ (see Eq. \ref{equ:kcoord}) are related to the $(x,y)$-coordinates in the object space by:
\begin{equation}\label{equ:trafo1}
(u,v) =(x f_s/f_F, y f_s/f_F).
\end{equation} 
For the sake of convenience the optical power of the Fourier transforming lenses within the following setups will be assigned to be  $f_F^{-1}=f_s^{-1}$, which just means that the variable optical power $m f_s^{-1}$ of the employed (combi-)saddle lenses is from now on measured in units ($m$) of the optical power of the Fourier transforming lenses.  In this case the scaling of the coordinates  $(u,v)$ in the plane of the optically performed image convolution corresponds to that of the original $(x,y)$-coordinates in the object space, according to Eq. \ref{equ:trafo1}. Furthermore the magnification of the saddle lens telescope then automatically corresponds to the factor $m$.
The optical Fourier transform of a saddle lens with a quadrupole optical power of $m f_s^{-1}$ (blue) now becomes (according to Eq. \ref{equ:FTSaddle}):
\begin{equation}\label{equ:saddlefft}
\mathcal{\mathrm{F}}\left[\exp(i \frac{m \pi}{\lambda f_s}  [x^2- y^2])\right] = \exp(-i  \frac{\pi}{m \lambda f_s} [x^2- y^2]).
\end{equation}
There, $\mathcal{\mathrm{F}}(...)$ denotes the optically performed Fourier operation according to Fig. \ref{fft}(a) or (b), using Fourier transforming lenses with a focal length of $f_s$. Thus a saddle lens with a quadrupole optical power $\pm m f_s^{-1}$ is transformed into a conjugate saddle lens with a quadrupole optical power $\mp f_s^{-1}/m$.
\begin{figure}[t]
\begin{center}
\includegraphics[width=\columnwidth]{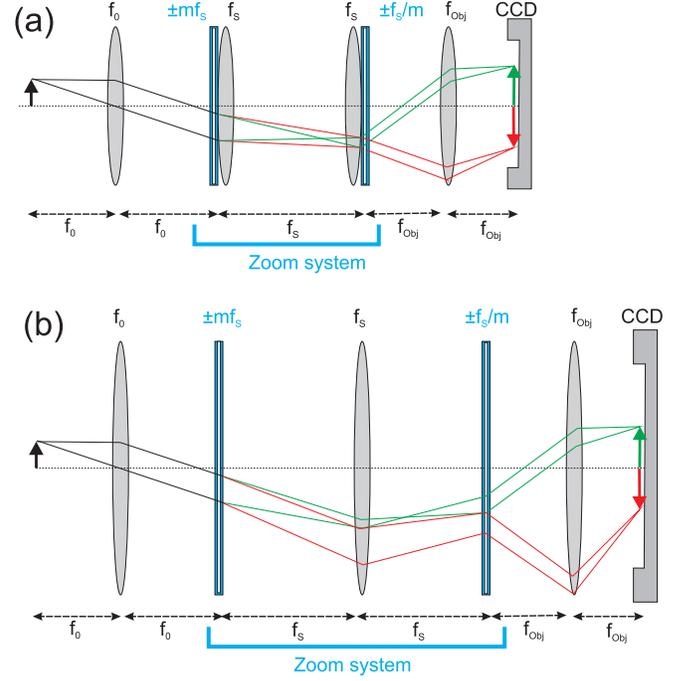}
\end{center}
\caption[]{Two examples for a zoom system using a combination of saddle lenses (blue) and spherical lenses (gray). Setups (a) and (b) are based on the optical Fourier transform setups described in  Fig. \ref{fft}(a) and (b), respectively. In both setups an object represented  by an arrow is imaged onto a camera. The focal lengths of the lenses and their distances are indicated in the figure. The red and green rays are actually propagating in orthogonal optical planes, namely in the $(y,z)-$plane (green), and in the $(x,z)-$plane (red). In the example the rays are sketched for the case $m=2$.}
\label{setup1}
\end{figure}

Furthermore the $(u,v)$-coordinates in Eq. (\ref{equ:result1}) can be replaced by $(x,y)$-coordinates, resulting in:
\begin{eqnarray}\label{equ:result2}
&& g^{\mathrm{R}}\left(\frac{x}{m},\frac{y}{m}\right) =\mathrm{e}^{-i \frac{\pi}{m \lambda f_s} [x^2- y^2]} \times \\ && \mathcal{\mathrm{F}}\left\{ \mathrm{e}^{-i \frac{m \pi}{\lambda f_s} [x^2-y^2]} \cdot \mathcal{\mathrm{F}}\left[\mathcal{\mathrm{F}}\{g(x,y)\} \cdot \mathrm{e}^{-i\frac{\pi}{m \lambda f_s}  [x^2- y^2]} \right]\right\}. \nonumber
\end{eqnarray}

A more concise version of this Equation is possible with the convention used in Eq. \ref{equ:saddle}, where a saddle lens is just represented by its quadrupole optical power. In this case Eq. \ref{equ:result2} may be abbreviated as:

\begin{equation}\label{equ:result3}
g^{\mathrm{R}}\left(\frac{x}{m},\frac{y}{m}\right) =S(\frac{f_s^{-1}}{m}) \mathcal{\mathrm{F}}\left\{ S(m f_s^{-1}) \cdot \mathcal{\mathrm{F}}\left[\mathcal{\mathrm{F}}\{g(x,y)\} \cdot S(\frac{f_s^{-1}}{m}) \right]\right\}.
\end{equation}

This equation can be interpreted as a sequence of operations which can be optically performed by any of the two setups shown in Fig. \ref{setup1}(a) and (b). A similar setup for performing generalized coordinate transforms has also been discussed in  \cite{Wang}, and investigated in \cite{Iemmi}, where the anamorphic lenses have been displayed on a spatial light modulator. In the Figure the object $g(x,y)$ to be imaged is indicated by a black arrow at the left. A first lens with focal length $f_0$ located at the distance $f_0$ performs a Fourier  transform $\mathcal{\mathrm{F}}\{g(x,y)\}$ of the object in the plane of the first combi-saddle lens (blue). The Fourier transformed image passes the first combi-saddle lens with a quadrupole optical power of $m^{-1} f_s^{-1}$, which yields (after multiplication of the corresponding transmission functions) $\mathcal{\mathrm{F}}\{g(x,y)\} \cdot S(m^{-1}f_s^{-1})$, which is the term in the square brackets in Eq. \ref{equ:result3}. This field is then Fourier transformed using any of the optical setups sketched in the previous Fig. \ref{fft}(a) or (b), respectively. Thus in front of the second saddle lens we obtain the field $\mathcal{\mathrm{F}}\left[\mathcal{\mathrm{F}}\{g(x,y)\} \cdot S(m^{-1}f_s^{-1}) \right]$. This field is then multiplied with the transmission function of the second combi-saddle lens  $S(m f_s^{-1})$, leading to the term in the curly brackets in Eq. \ref{equ:result3}. A final Fourier transform of this field is performed by an objective lens with focal length $f_{\mathrm{Obj}}$ at a distance of $f_{\mathrm{Obj}}$ behind the second saddle lens. The Fourier transform is projected at the camera plane. Thus, up to a phase factor of a saddle lens $S(m^{-1}f_s^{-1})$ (corresponding to the first term in Eq. \ref{equ:result3}) the field in the camera plane corresponds to the right side of Eq. \ref{equ:result3}, yielding as a result the scaled and mirrored image $g^{\mathrm{R}}\left(x/m, y/m \right)$. If required, the phase of the image can be optically corrected by placing a third saddle lens, which acts as a field lens, with the conjugate transmission function of $S(-m^{-1}f_s^{-1})$ (which corresponds to a 90$^{\circ}$ rotated version of $S(+m^{-1}f_s^{-1})$) directly in the camera plane, or in an intermediate image plane of an expanded optical setup. The adjustment of the quadrupole optical powers of the included combi-saddle lenses for a certain magnification factor $m$ can then be performed by a symmetric rotation of their components into opposite directions by an angle $\alpha$ according to Eq. \ref{equ:factm1}, which keeps the main axes of the two combi-saddle lenses aligned. An additional rotation of all included (combi-)saddle lenses by an angle $\theta$ into the same direction leads to an adjustable image rotation of $2 \theta$. 

Figure \ref{setup1} indicates the optical rays for the two orthogonal main planes of the anamorphic system ($(y,z)-$plane in green, $(x,z)-$plane in red) for the case $m=2$. In one of the planes (green) the zoom system corresponds to a Kepler telescope, whereas it corresponds to a Galilei telescope in the other plane (red). The magnifications of the Kepler and Galilei telescopes are identical for all adjustable magnification factors $m$.

Fig. \ref{setup1} also shows that the actual zoom system consists of the optical setup between the two saddle lenses (indicated in the figure), which acts as an afocal telescope. A condition for the correct operation of the zoom system is that the light field emerging from any point of the object plane is transformed into a parallel beam before entering the telescope. This condition can also be fulfilled in a compressed setup in which the first spherical lens is located directly in front of the first saddle lens. Similarly, the last (objective) lens can also be placed directly behind second saddle lens.

\begin{figure}[t]
\begin{center}
\includegraphics[width=\columnwidth]{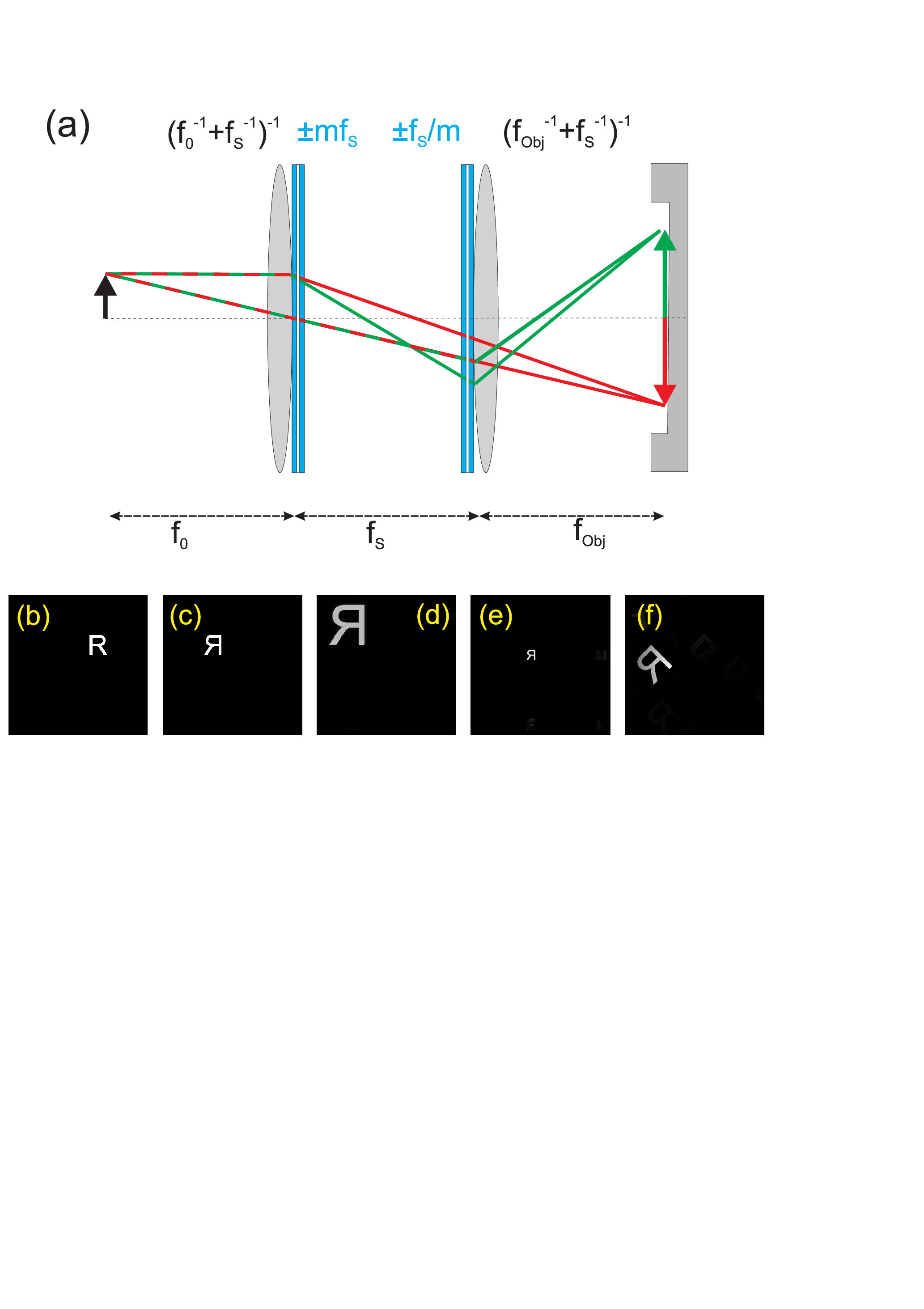}
\end{center}
\caption[]{(a): More compact zoom system with saddle lenses based on the layout of Fig.\ref{setup1}(a). Now the spherical front lens (optical power $f_0^{-1}$) and the first Fourier transforming lens (optical power $f_s^{-1}$) are combined into a single spherical lens with optical power of $f_0^{-1}+f_s^{-1}$, which is placed directly in front of the first combi-saddle lens. Analogously the second Fourier transforming lens and the objective lens are combined into a single spherical lens with an optical power of $f_s^{-1}+f_{Obj}^{-1}$, which is placed directly behind the second combi-saddle lens. (b-f) show wave-optical simulations of the imaging properties of the setup. (b): Test object to be imaged. (c): Resulting image in CCD plane for $m=1$. (d): Resulting image for $m=2$. (e): Resulting image for $m=0.5$. (f): Resulting image for $m=1.5$, and an additional rotation of both of the saddle lenses by an angle of $\theta=22.5 ^{\circ}$.}
\label{setup2}
\end{figure}

This setup  can still be further compressed by combining all adjacent spherical lenses into single spherical lenses with correspondingly adapted optical powers. Such a compressed setup based on the expanded setup of Fig. \ref{setup1}(a) is sketched in Fig. \ref{setup2}(a), with the corresponding focal lengths of the lenses indicated in the figure. A simulation of its imaging properties is shown in Fig. \ref{setup2}(b-f). The simulation is based on wave optical propagation of the input amplitude object shown in Fig. \ref{setup2}(b) through the optical system, using the spectrum of plane wave propagation method and assuming thin optical elements which are purely characterized by their transmission function. Figure \ref{setup2}(c) shows the image in the camera plane for the case $m=1$. As expected, the image is mirrored with respect to the $y-$axis. In  Fig. \ref{setup2}(d) the same simulation is performed for $m=2$, resulting in an image magnification by this factor. In (e) a factor $m=0.5$ was chosen, resulting in a corresponding image demagnification. Finally in (f) a factor $m=1.5$ was chosen, and additionally the two combi-saddle lenses were rotated by the same angle $\theta = 22.5^{\circ}$. This results in an image magnification by the factor $m=1.5$, and additionally in a rotation of the whole image by the angle $2 \theta = 45^{\circ}$. 

In practice the setup of Fig. \ref{setup2}(a) can be even more compressed if the optical powers of the two remaining spherical lenses are incorporated into the respective adjacent combi-saddle lens elements, for example by adding spherical lens terms with half of the required optical power to each of the two individual saddle lenses of a combi-saddle lens. The resulting zoomable telescope then contains no spherical lenses any more, but just consists of four generalized toroidal lenses combined into two pairs of tunable combi-toroidal lenses, respectively.

\section{Conclusion}
Using combi-saddle lenses in a novel telescope setup, which combines a Kepler- and a Galilei telescope in two orthogonal planes, one obtains a continuously tunable zoom system, with a magnification which is purely controlled by rotating the saddle lenses. The combi-saddle lenses may be realized as diffractive optical elements, which has the advantage that the elements are very thin, and geometrical optical aberrations due to the thickness of implemented glass elements are minimized. On the other hand, the elements can also be realized as refractive elements, with the advantage of having full efficiency independent of the wavelength, and without the high dispersion of diffractive optical elements. By using a third combi-saddle lens as a field lens in the image plane of the camera system (or in a conjugate intermediate image plane), the optical system becomes phase preserving, i.e. the phase in the image plane corresponds to that in the object plane.

The principle of the method depends on the features of saddle lenses that their Fourier transform is again a saddle lens, and that a convolution of an input image with a saddle lens basically corresponds to a Fourier transform of the input image itself, which is however scaled by a factor depending on the quadrupole optical power of the saddle lens. Actually, standard (parabolic and cylindrical) lenses have analogous features, but a saddle lens is the only one of these lens types which is tunable by a relative rotation of two adjacent lens elements. In a ray-optical description a saddle lens telescope corresponds to a combined Kepler and Galilei telescope in two orthogonal planes, where both of the two telescope types automatically have the same magnification for all adjusted zoom factors $m$. The optical system has the feature that it produces a mirror image of the object with respect to a selectable axis, which allows one to rotate the image by an angle $2\theta$ by rotating all implemented saddle lenses by the same angle of $\theta$ into the same direction.

An afocal version of the tunable telescope as sketched in Fig. \ref{classical}(c) may find applications as a variable beam expander, e.g. in laser material processing, or as a tunable eyepiece in microscopes, telescopes, or spotting scopes. With an additional focusing unit and/or an additional imaging lens the system may be also used as a zoomable camera objective, or a zoomable image (or video) projection device. The feature of the telescope to produce a mirrored image can be corrected by inserting an additional mirror at any position in the beam path, e.g. by a side-view system. There the feature that the image can be continuously rotated allows one to obtain an erect image even when "looking" from different sides into the telescope.   
  
Interestingly, the wave-optical simulation by processing Eq. (\ref{equ:result3}) suggests a numerical method to rotate and scale an image, which does not require numerical interpolation in the image space, but is purely based on a sequence of three Fourier transformations, and point wise array multiplications with analytically rotatable saddle lens transmission functions. This might be particularly interesting for numerically rotating and scaling of complex images with strongly modulated phase functions, like phase-wrapped interferograms, since interpolation of such erratic complex functions is typically quite challenging \cite{Lohmann, Unser, Larkin}.

\section{Funding Information}
Austrian research funding association (FFG), contract No. 864729.



\end{document}